\documentclass[
aip,apl,
amsmath,amssymb,
reprint]{revtex4-1}

\usepackage{graphicx}
\usepackage{units}

\begin{document}


\title{Intermodulation Electrostatic Force Microscopy for imaging Surface Photo-Voltage}

\author{Riccardo Borgani}
\email{borgani@kth.se}
\author{Daniel Forchheimer}
\affiliation{Nanostructure Physics, Royal Institute of Technology, 10691 Stockholm, Sweden}
\author{Jonas Bergqvist}
\affiliation{\mbox{Department of Physics, Chemistry and Biology, Link\"oping University, 58183 Link\"oping, Sweden}}
\author{Per-Anders Thor\'en}
\affiliation{Nanostructure Physics, Royal Institute of Technology, 10691 Stockholm, Sweden}
\author{Olle Ingan\"as}
\affiliation{\mbox{Department of Physics, Chemistry and Biology, Link\"oping University, 58183 Link\"oping, Sweden}}
\author{David B. Haviland}
\email{haviland@kth.se}
\affiliation{Nanostructure Physics, Royal Institute of Technology, 10691 Stockholm, Sweden}

\date{\today}

\begin{abstract}
We demonstrate an alternative to Kelvin Probe Force Microscopy (KPFM) for imaging surface potential. The open-loop, single-pass technique applies a low-frequency AC voltage to the atomic force microscopy tip while driving the cantilever near its resonance frequency. Frequency mixing due to the nonlinear capacitance gives intermodulation products of the two drive frequencies near the cantilever resonance, where they are measured with high signal to noise ratio. Analysis of this intermodulation response allows for quantitative reconstruction of the contact potential difference.  We derive the theory of the method, validate it with numerical simulation and a control experiment, and we demonstrate its utility for fast imaging of the surface photo-voltage on an organic photo-voltaic material.
\end{abstract}

\maketitle


One of the most popular and useful methods of Electrostatic Force Microscopy (EFM) is Kelvin Probe Force Microscopy (KPFM)\cite{Nonnenmacher1991} which provides a measurement of the contact potential difference $V_{\mathrm{CPD}}$ (sometimes referred to as the surface potential). KPFM is widely used for advanced imaging of composite polymeric materials\cite{Cadena2013} and for imaging of the local work function on the surface of organic photo-voltaic materials\cite{Hoppe2005}. Although KPFM is a useful technique to investigate electric properties of surfaces at the nanoscale, the signal-to-noise ratio, accuracy and speed are limited by the additional feed-back loops commonly used in its implementations\cite{Collins2013}. To overcome these limitations, an open-loop technique was first proposed by Takeuchi et al.\cite{Takeuchi2007} to image the contact potential difference in vacuum. Later the technique was used to measure the potential of nanoparticles in liquid\cite{Kobayashi2011} and to characterise ferroelectric thin films\cite{Collins2012}.

In this paper we propose and demonstrate an open-loop technique that exploits the intermodulation (frequency mixing) of an electrostatic drive force and a mechanical drive force, to up-convert the electrostatic frequency to the first flexural resonance where the high quality factor allows for a more sensitive measurement. The contact potential difference can be imaged in a single-pass, allowing for imaging times shorter than $5$ min with $256\times256$ pixel resolution.


The electrostatic energy stored in a system of two perfect conductors is $E_{\text{EL}}=\frac{1}{2}CV^{2}$, where~$C$ is the capacitance and~$V$ the electrostatic potential difference between the two. The attractive electrostatic force is therefore;

\begin{equation}
F_{\mathrm{EL}}=\frac{1}{2}\frac{\mathrm{\partial}C}{\mathrm{\partial}z}V^{2} \label{eq:F}
\end{equation}

\noindent where~$z$ is the distance between the two conductors. In EFM the two conductors are the conductive tip and the sample substrate, which can be approximated as an axially symmetric electrode and an infinite conducting plane respectively. The resulting capacitance gradient varies as a non-linear function of~$z$ that depends on the tip geometry\cite{Hudlet1998}.

Intermodulation EFM (ImEFM) excites the cantilever with a shaker piezo at frequency~$\omega_{\mathrm{D}}$ close to resonance~$\omega_{0}$, while at the same time an AC voltage is applied to the cantilever at frequency~$\omega_{\mathrm{E}}\ll\omega_{\mathrm{D}}$. The total potential between the tip and the sample is:
\begin{equation}
V\left(t\right)=V_{\mathrm{CPD}}+V_{\mathrm{AC}}\cos\left(\omega_{\mathrm{E}}t+\phi_\mathrm{E}\right)
\label{eq:V}
\end{equation}
where $V_{\mathrm{CPD}}$ is the contact potential difference between the tip and the sample, assumed to be function of the in-plane tip position, and $\phi_\mathrm{E}$ is an arbitrary phase delay between the applied voltage and the lock-in reference signal. For a high $Q$ cantilever oscillation the tip motion is dominantly harmonic at $\omega_{\mathrm{D}}\approx \omega_0$. The time evolution of the tip-sample distance may be written,
\begin{equation}
z\left(t\right)\approx h+A_{\mathrm{D}}\cos\left(\omega_{\mathrm{D}}t+\phi_\mathrm{D}\right)
\label{eq:z}
\end{equation}
where $h$ is the tip rest position (or static probe height), and $A_{\mathrm{D}}$ and $\phi_\mathrm{D}$ are the oscillation amplitude and phase which depend on the drive force and on the interaction with the surface.

The capacitance gradient is a non-linear function of the tip-sample separation~$z$. We define~$C'=\frac{\mathrm{\partial}C}{\mathrm{\partial}z}$ and perform a polynomial expansion around the resting position~$h$:

\begin{equation}
C'\left(z\right)=\sum\limits_{n=0}^{+\infty}\frac{1}{n!}\left.\frac{\mathrm{\partial}^nC'}{\mathrm{\partial}z^n}\right|_{h}\left(z-h\right)^n
\label{eq:Cz_taylor}
\end{equation}

\noindent Which together with (\ref{eq:z}) gives:

\begin{eqnarray}
C'\left(t\right) & = & \sum\limits_{n=0}^{+\infty}\frac{A_\mathrm{D}^n}{n!}\left.\frac{\mathrm{\partial}^nC'}{\mathrm{\partial}z^n}\right|_{h}\cos^n\left(\omega_\mathrm{D}t+\phi_\mathrm{D}\right)\nonumber\\*
 & = & \sum\limits_{k=0}^{+\infty}a_k\cos\left(k\omega_\mathrm{D}t+k\phi_\mathrm{D}\right)
\label{eq:Cz_harmonics}
\end{eqnarray}

\noindent where the coefficients $a_k$ are linear combinations of the terms $C'^{\left(n\right)}A_{\mathrm{D}}^n/n! $. Inserting Eq.s~(\ref{eq:V}) and~(\ref{eq:Cz_harmonics}) into (\ref{eq:F}) gives:

\begin{eqnarray}
F_{\mathrm{EL}} & = & \frac{1}{2}C'\left(t\right)\left[2V_{\mathrm{CPD}}V_{\mathrm{AC}}\cos\left(\omega_{\mathrm{E}}t+\phi_\mathrm{E}\right)\right.\nonumber\\*
 & + & \left.\frac{1}{2}V_{\mathrm{AC}}^{2}\cos\left(2\omega_{\mathrm{E}}t+2\phi_\mathrm{E}\right)+\frac{V_{\mathrm{AC}}^{2}}{2}+V_{\mathrm{CPD}}^{2}\right]
\end{eqnarray}

\noindent Re-arranging terms, it is possible to separate the electrostatic force in components at different frequencies~$\omega_{i}$ with complex amplitudes~$\hat{F}_{\omega_{i}}$:

\begin{subequations}
\begin{eqnarray}
\hat{F}_{\mathrm{DC}} & = & \frac{1}{2}a_{0}\left(\frac{V_{\mathrm{AC}}^{2}}{2}+V_{\mathrm{CPD}}^{2}\right)
\\
\hat{F}_{\omega_{\mathrm{D}}} & = & \frac{1}{2}a_{1}\left(\frac{V_{\mathrm{AC}}^{2}}{2}+V_{\mathrm{CPD}}^{2}\right)\mathrm{e}^{\mathrm{i}\phi_\mathrm{D}}
\\
\hat{F}_{\omega_{\mathrm{E}}} & = & a_{0}V_{\mathrm{CPD}}V_{\mathrm{AC}}\mathrm{e}^{\mathrm{i}\phi_\mathrm{E}}
\label{eq:Fwe}
\\
\hat{F}_{2\omega_{\mathrm{E}}} & = & \frac{1}{4}a_{0}V_{\mathrm{AC}}^{2}\mathrm{e}^{\mathrm{i}2\phi_\mathrm{E}}
\label{eq:F2we}
\\
\hat{F}_{\omega_{\mathrm{D}}\pm\omega_{\mathrm{E}}} & = & \frac{1}{2}a_{1}V_{\mathrm{CPD}}V_{\mathrm{AC}}\mathrm{e}^{\mathrm{i}\phi_\mathrm{D}}\mathrm{e}^{\pm\mathrm{i}\phi_\mathrm{E}}
\label{eq:Fwdwe}
\\
\hat{F}_{\omega_{\mathrm{D}}\pm2\omega_{\mathrm{E}}} & = & \frac{1}{8}a_{1}V_{\mathrm{AC}}^{2}\mathrm{e}^{\mathrm{i}\phi_\mathrm{D}}\mathrm{e}^{\pm\mathrm{i}2\phi_\mathrm{E}}
\label{eq:Fwd2we}
\end{eqnarray}
\end{subequations}

\noindent Other force components are present at frequencies $k\omega_{\mathrm{D}}$, $k\omega_{\mathrm{D}} \pm \omega_{\mathrm{E}}$ and $k\omega_{\mathrm{D}} \pm 2\omega_{\mathrm{E}}$. However in this analysis we limit ourselves to the components at low frequency and around the cantilever drive frequency since they are the ones experimentally detectable with good signal-to-noise ratio.

With the driving scheme used in ImEFM, it is possible to extract $V_{\mathrm{CPD}}$ from the measurement of the force components at low frequency~(\ref{eq:Fwe})-(\ref{eq:F2we}):

\begin{equation}
V_{\mathrm{CPD}}=\frac{V_{\mathrm{AC}}}{4}\frac{\hat{F}_{\omega_{\mathrm{E}}}}{\hat{F}_{2\omega_{\mathrm{E}}}}\mathrm{e}^{\mathrm{i}\phi_\mathrm{E}}\label{eq:Vcpd_low}
\end{equation}

or from the components at high frequency~(\ref{eq:Fwdwe})-(\ref{eq:Fwd2we}):

\begin{equation}
V_{\mathrm{CPD}}=\frac{V_{\mathrm{AC}}}{4}\frac{\hat{F}_{\omega_{\mathrm{D}}\pm\omega_{\mathrm{E}}}}{\hat{F}_{\omega_{\mathrm{D}}\pm2\omega_{\mathrm{E}}}}\mathrm{e}^{\pm\mathrm{i}\phi_\mathrm{E}}\label{eq:Vcpd_imp}
\end{equation}

\noindent The phase factor $\phi_\mathrm{E}$ can be set to zero by ensuring that the AC voltage is in phase with the lock-in reference signal.

Obtaining $V_{\mathrm{CPD}}$ from Eq.~(\ref{eq:Vcpd_low}) or from~(\ref{eq:Vcpd_imp}) is in principle equivalent, however in experimental conditions noise is present in the detection system and the cantilever resonance allows for a measurement of the force components~(\ref{eq:Fwdwe}) and~(\ref{eq:Fwd2we}) with much higher signal-to-noise ratio, limited in sensitivity only by the thermal noise force. Note that $V_{\mathrm{CPD}}$ depends on a measurement of force, which is obtained from the cantilever motion by a calibration procedure\cite{Sader1999,Higgins2006}. However in ImEFM $V_{\mathrm{CPD}}$ is proportional to the ratio of two forces. Thus only the frequency dependence of the cantilever transfer function (resonance frequency and quality factor) are significant to the calibration while the mode stiffness and the optical lever responsivity fall out of the ratio.

The expressions for the contact potential difference hold for any form of the capacitance gradient since we did not truncate the polynomial expansion (\ref{eq:Cz_taylor}) to a finite order. In particular, the validity of the technique does not depend upon the assumption that the capacitance gradient, or its first derivative, is constant in the oscillation range\cite{Maragliano2014}. Under the condition of low drive amplitude, it is however possible to approximate the capacitance gradient as a linear function of $z$, i.e. truncate expansion (\ref{eq:Cz_taylor}) at $n=1$. Eq. (\ref{eq:Cz_harmonics}) gives $a_{0}=C'$ and $a_{1}=A_{\mathrm{D}}C''$, and it is possible to evaluate the capacitance gradient and its first derivative at $h$ from the measured force components:

\begin{subequations}
\begin{eqnarray}
C' & = & 4\frac{\hat{F}_{2\omega_{\mathrm{E}}}}{V_{\mathrm{AC}}^{2}}\mathrm{e}^{-\mathrm{i}2\phi_\mathrm{E}}\label{eq:Cz_exp}\\
C'' & = & 8\frac{\hat{F}_{\omega_{\mathrm{D}}\pm2\omega_{\mathrm{E}}}}{A_{\mathrm{D}}V_{\mathrm{AC}}^{2}}\mathrm{e}^{-\mathrm{i}\phi_\mathrm{D}}\mathrm{e}^{\mp\mathrm{i}2\phi_\mathrm{E}}\label{eq:Czz_exp}
\end{eqnarray}
\end{subequations}


We simulated ImEFM by numerically integrating the differential equation that models the cantilever dynamics:

\begin{equation}
\ddot{d}+\frac{\omega_{0}}{Q}\dot{d}+\omega_{0}^{2}d=\frac{\omega_{0}^{2}}{k}F_{\mathrm{TOT}}\left(t,z\right)
\end{equation}

where $d=z-h$ is the cantilever deflection, $z$ is the tip-sample distance, $h$ is the tip resting position, and $\omega_{0}$, $Q$ and $k$ are the resonance frequency, quality factor and stiffness of the first flexural mode of the cantilever. The total force on the cantilever $F_{\mathrm{TOT}}$ is given by three contributions: a sinusoidal drive force close to the first flexural resonance due to the inertial actuation, the electrostatic force for an axial symmetric electrode over an infinite plane surface\cite{Hudlet1998}, and the tip-surface force modelled by a Lennard-Jones potential\cite{14934}. The result of the numerical integration is the cantilever deflection signal as a function of time. We compute the Discrete Fourier Transform to obtain the frequency spectrum of cantilever deflection $\hat{d}\left(\omega\right)$, from which we then calculate the force spectrum $\hat{F}\left(\omega\right)$ by multiplying with the cantilever inverse response function $\hat{\chi}^{-1}$:

\begin{subequations}
\begin{eqnarray}
\hat{F}\left(\omega\right) & = & \hat{\chi}^{-1}\hat{d}\left(\omega\right)\\
\hat{\chi}^{-1}\left(\omega\right) & = & k\left(1+\mathrm{i}\frac{\omega}{\omega_{0}Q}-\frac{\omega^{2}}{\omega_{0}^{2}}\right)
\end{eqnarray}
\end{subequations}

\noindent We finally calculate $V_{\mathrm{CPD}}$ according to Eq.s (\ref{eq:Vcpd_low}) and (\ref{eq:Vcpd_imp}).

\begin{figure}
\includegraphics[width=\columnwidth]{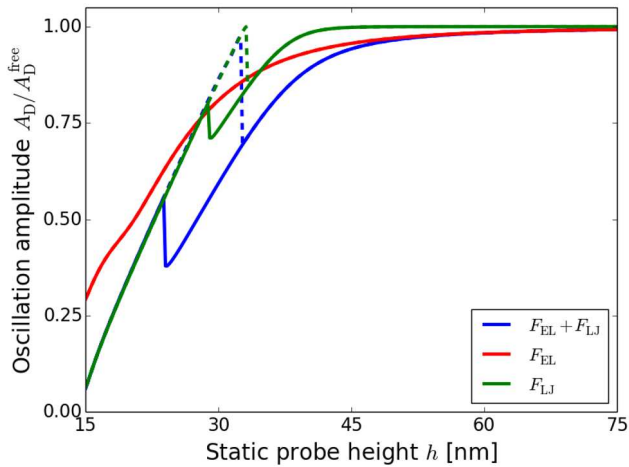}
\caption{\label{fig:simulation}Simulated cantilever oscillation amplitude at the mechanical drive frequency as a function of distance from the sample during approach (solid line) and retract (dashed line) with initial amplitude $A_\mathrm{D}^\mathrm{free}=20~\mathrm{nm}$. The AC electrostatic force causes the amplitude to drop faster than in the case with short-range forces only. Hysteresis is visible in the approach-retract curves with short-range interactions.}
\end{figure}

From the results of the simulation it is possible to investigate the validity of assumption~(\ref{eq:z}). When the probe softly interacts with the sample the motion is prevalently harmonic, i.e. the amplitude component at the drive frequency is higher than any other component by two or more orders of magnitude. The numerical integrator can compute all the intermodulation products, however, when a realistic detector noise level is added to the simulation, only two intermodulation products are visible above the noise on each side of the drive frequency. On the other hand, when the interaction with the surface is stronger more peaks arise above the noise level and the reconstruction of the $V_{\mathrm{CPD}}$ is not accurate.

By simulating the cantilever dynamics we investigate the effect of the electrostatic force and the tip-surface force (modelled by a Lennard-Jones potential) separately. The simulations confirm that it is the non-linear electrostatic force that up-converts the electrostatic frequency to the first flexural resonance. The short range tip-surface force is not required to measure the intermodulation spectrum. Moreover FIG.~\ref{fig:simulation} shows that by adding the AC electrostatic force to the tip-surface interaction, more frequency components are available to accommodate the oscillation energy of the cantilever, causing the oscillation amplitude to drop at larger distance from the surface and thus giving a more stable feed-back in the so-called non-contact regime where attractive forces dominate over repulsive forces. It is in this regime that our single pass scan is performed, corresponding to a static probe height $h \approx 60 \mathrm{nm}$ (for a free oscillation amplitude $A_\mathrm{D} = 30 \mathrm{nm}$ and an amplitude set-point of 90\%).


Experiments were performed on a JPK NanoWizard 3 AFM mounted on an inverted optical microscope. The generation of the electrical and mechanical drive signals and the acquisition of the intermodulation spectra were performed with an intermodulation lock-in analyser\cite{Tholen2011}.
\begin{figure}
\includegraphics[width=\columnwidth]{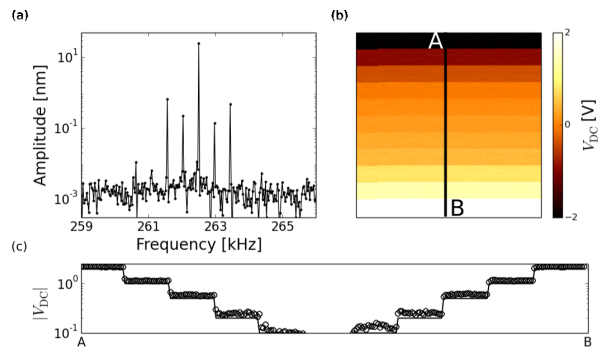}
\caption{\label{fig:apply_dc}Typical intermodulation spectrum around resonance when performing ImEFM (a). As the cantilever scans over a gold surface (b), different steps of DC potential are applied to the sample (c, solid line). ImEFM is able to measure the correct variation in surface potential.}
\end{figure}
To experimentally validate the technique, we applied a series of DC potential steps with different amplitudes while performing ImEFM on a gold substrate (FIG.~\ref{fig:apply_dc}b). We used a Cr-Au coated cantilever by Mikromash with $300.5$ kHz resonance frequency, driven close to resonance with a free oscillation amplitude of $35$ nm. We applied a $6$ V AC potential at $469$ Hz with a pixel time of $2.1$~ms. The technique was able to measure the intermodulation spectrum (FIG.~\ref{fig:apply_dc}a) and reconstruct the potential applied to the sample within a few \% and with very low noise (FIG.~\ref{fig:apply_dc}c).

\begin{figure*}
\includegraphics[width=2\columnwidth]{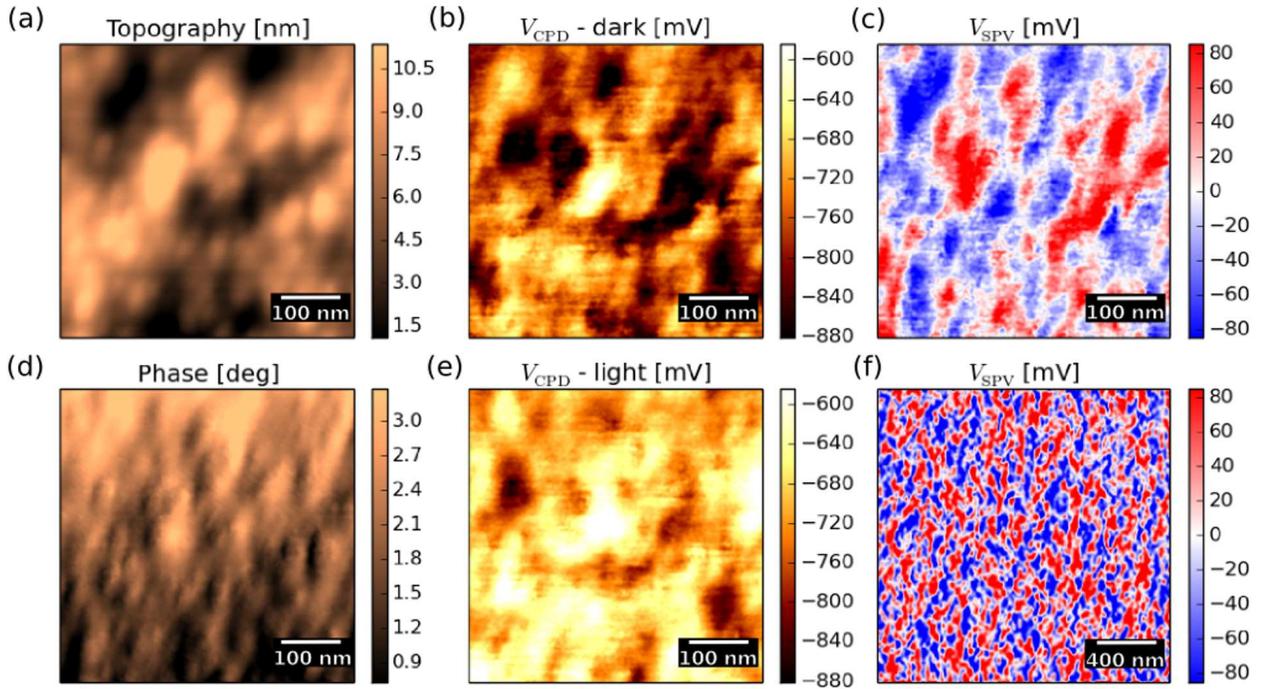}
\caption{\label{fig:solarcell}(a)-(e) ImEFM on a TQ1:PCBM:C60 sample, $500$ nm scan size with $256\times256$ pixels resolution. The total acquisition time is $5$ min. Despite the very flat topography and limited contrast in the phase image, different domains are clearly visible in the contact potential difference images in dark and under illumination, and especially in the surface photo-voltage image which shows domains with size of order $50$-$100$ nm. The domains appear to be regular across the surface as shown with a bigger scan size of $2 \mathrm{\mu m}$ (f).}
\end{figure*}

We apply the technique to spatially resolve the photo-generation of charge in a TQ1:PCBM:C60\cite{Lindqvist2014,Lindqvist2014-2,DiazdeZerioMendaza2014} thin film spin-coated on an ITO electrode. We acquire two $V_{\mathrm{CPD}}$ images, one in dark and one under illumination, during the scan trace and re-trace respectively. We then calculate the surface photo-voltage $V_{\mathrm{SPV}}$ as the difference of the trace and re-trace $V_{\mathrm{CPD}}$ images\cite{Spadafora2010}. FIG.~\ref{fig:solarcell} highlights the presence of domains with size of the order of $50$ nm, and we notice a correlation between areas of low work function (low $V_{\mathrm{CPD}}$ in the dark), and areas of high surface photo-voltage. Areas with high values of $V_{\mathrm{SPV}}$ correspond to an increased $V_{\mathrm{CPD}}$ under illumination, which can be explained by a higher concentration of photo-generated holes than electrons, and therefore a region with a high concentration of donors (lower work function).


We demonstrated an EFM technique for mapping surface potential with high signal-to-noise ratio, making use of the high force sensitivity of the cantilever mechanical resonance and frequency mixing due to the nonlinear capacitance gradient. Being an open-loop technique, the feed-back induced cross talk is avoided and the measurement speed is not limited by the feed-back bandwidth. The absence of an applied DC bias makes this technique good for characterising bias sensitive systems and materials with high work function that would require additional voltage amplifiers with feed-back based techniques. Finally, the ability to perform a single-pass measurement significantly lowers the imaging time and provides higher lateral resolution than interleaved lift-mode techniques.

\begin{acknowledgments}
The authors acknowledge financial support from the Swedish Research Council (VR), the Knut and Alice Wallenberg Foundation, and the Olle Engkvist Foundation. We are grateful for fruitful discussions with Liam Collins.
\end{acknowledgments}

\bibliography{my_bib}

\end{document}